# Phononic engineering with nanostructures for hot carrier solar cells

Jean-Francois Guillemoles[1,2], Gavin Conibeer[1], Martin Green[1]

[1]Centre of Excellence for Advanced Silicon Photovoltaics and Photonics, University of New South Wales, Sydney 2052 Australia,
email: g.conibeer@unsw.edu.au , m.green@unsw.edu.au

[2]On leave from IRDEP, UMR CNRS-ENSCP-EDF 7174, Paris, email: jf-guillemoles@enscp.fr

**Abstract**: Hot Carrier solar cells have long been recognized as an attractive contender in the search for high efficiency photovoltaic devices but their fabrication requires solution of two important material challenges: finding materials with drastically reduced carrier cooling rates and realization of selective energy contacts to extract the photogenerated carriers. This paper is concerned with the problem of absorber phononic engineering to reduce carrier cooling rates. The physics of carrier cooling is explored and experimental data of other authors are discussed with a view to assessing the potential of state-of-the art nanostructured materials for PV conversion. A tentative initial calculation based on the thermalisation in these nanostructures and assuming all other aspects as ideal, gives 54% efficiency at 2500suns as compared to 54% for no thermalisation at all. Phononic band gap engineering to further reduce carrier cooling or bring down the threshold concentration is discussed.

## 1 Hot carrier cell: challenges

In the pursuit of practical devices with a very high photovoltaic efficiency, the hot carrier cell is a route displaying a number of advantages. Under full concentration, in the radiative limit, above 86% efficiency could be achieved in principle, whereas under global AM1.5 the limit is above 67% [1]. Very high efficiencies can be achieved with a single absorber of $E_g$ lower than 1 eV. For an ideal device there is no spectral sensitivity and even for non-ideal devices the sensitivity is small. Furthermore the overall fabrication could be fairly simple, as compared to tandem cells, which makes it a very attractive route to high efficiencies.

The principles of operation are illustrated in Fig. 1. The concept of the hot carrier cell is to collect photo-generated carriers before they have time to thermalise to the band edge [1-3]. The concept allows for thermalisation within the carrier populations themselves – a time period on the order of a picosecond - but not with the lattice.[^1] The absorber has a hot carrier distribution at temp $T_H$. Carriers cool isoentropically in the mono-energetic contacts to $T_A$: their kinetic energy being converted into useable potential energy. [2] The difference of the Fermi levels of these two contacts manifests as a difference in chemical potential of the carriers at each contact and hence an external voltage, V.

The challenges to produce such devices fall into 2 categories: (i) keeping carriers hot without heating the lattice (at least not beyond the selective contacts) and (ii) achieving a fast extraction of hot carriers through a narrow allowed energy range. [2] While these challenges are tough, it seems they can be met using newly available materials and more specifically nanostructured semiconductors, such as those based on III-V semiconductors in [4,5]. At UNSW, theoretical work on slowed cooling in nanostructures [6,7] has indicated some appropriate mechanisms, with some preliminary time-resolved photoluminescence evidence of prolonged occupation of multiple energy levels [8]. Work on selective energy contacts based on Si QD nanostructures has achieved some experimental success [9,10].

This paper is concerned with carrier cooling rates and absorber phononic engineering to reduce carrier cooling rates.

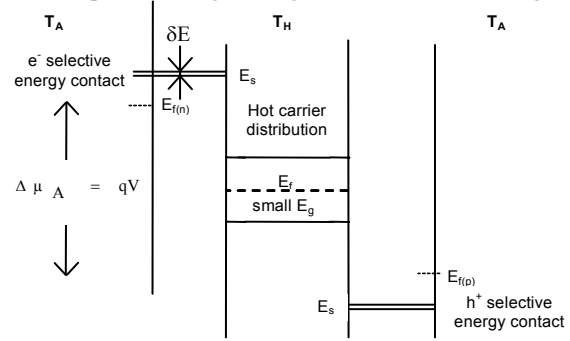

**Figure 1:** Hot carrier cell. All carriers leave the upper band at a common energy and are returned to the lower band at a common energy using energy selective contacts.

## 2 Carrier cooling

Carrier cooling in semiconductors and nanostructures has generated interest in the past decades following the realization that low dimensional semiconducting structures (LDS) do display significantly reduced hot carrier cooling rates [4,5].

Reasons for these reduced cooling rates in LDS are:
- more stringent conservation rules in the carrier-phonon interaction that couples carriers to fewer vibrational modes;
- carrier localization that prevents carrier cooling by out diffusion from the hot phonon region [11], a mechanism very effective on short time scales. Both lead to efficient reheating of carriers by emitted optical phonons (equivalent to renormalization of the free carrier calorific capacity).

Thermalization occurs via either the (relatively slow) production of weakly coupled, off zone centre, optical phonons or via the decay of zone centre LO phonons into 2LA phonons via anharmonic interactions (where the rate is dependent on the availability of states for these interactions) [4,5,12].

Many data have been made available in the literature in the past decades. In III-V QW especially, very detailed studies

[^1]: However, some efficient carrier-carrier thermalisation must be involved device operation, so that renormalisation of carrier energies can re-occupy energies which are denuded by extraction through the selective energy contacts.

have been undertaken [4,5,13]. The main lessons are that:
- using QW enables an increase in cooling times by about a factor of 100 for a given injection level as compared to bulk;
- the characteristic cooling time is roughly proportional to injection level (see Fig. 2) [13,14].

Another way of putting this is that the total heat transfer to the lattice is actually independent of the injection level, if the latter is high enough. Indeed, examination of experimental data from [14] shows that the characteristic cooling time $\tau$ vs carrier temperatures T follows an empirical law given by:

$$\tau \approx \frac{K.n}{T_h - T_a}$$

Where n is the injection level and $T_A$ is the ambient temperature and K, a material dependent parameter, can be related to the number of optical modes effectively coupled to the carriers. Following the definition of $\tau$, the heat lost by the electronic system to the lattice is given by:

$$\frac{dQ}{dt} = -n.\frac{dE}{dt} = \frac{\hbar\omega_{LO}}{K}.\exp\left(-\frac{\hbar\omega_{LO}}{kT_h}\right).(T_h - T_a)$$

With the above empirical model and assuming ideal selective contacts and optimal extraction, it is possible to infer maximal work that can be extracted from the cell taking into account the computed thermal losses. Preliminary computations show that at the illumination level used for the experiments (equivalent to about 2500 suns) efficiencies of **50%** could be achievable. This is in spite of thermal losses in state of the art materials (which have a non optimal band gap of 1.5 eV), as compared to **54 %** without such losses.[2] Hence, at these concentrations the loss to thermalisation for existing materials is not large. However, as operation at lower concentrations would be preferable and because of the difficulties in the fabrication of optimal contacts, it is desirable to search for further reduction in carrier cooling and/or a reduction in the threshold injection levels at which this occurs.

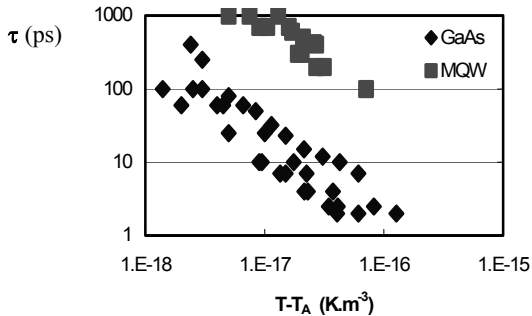

**Figure 2**: Relaxation times vs. T-$T_A$ scaled by carrier injection level for bulk and MQW samples. Data from [14,15].

## 3 Phononic band gap materials

Favorable conditions for a reduced LO phonon generation include polar materials (because of the Frohlich interaction ) and as shown in Fig. 2, nanostructuring a material can give a specific reduced cooling due to electron and phonon confinement effects [13]. Q dots should enhance this effect further, but the actual gain is difficult to assess from experimental data because hot carrier temperatures are not experimentally accessible.

Alternatively the LO->2LA decay could also be strongly reduced in phononic band gaps materials, such as mass-contrasted compounds such as InN. [6,7] Or, such gaps in the phonon dispersion can open up in the presence of interfaces in nanostrucured materials with an acoustic mismatch. For QDs there can be gaps in LA DOS that are complete in k-space, and, if suitably located at half the energy of the optical phonon at zone centre, these could prevent the LO decay. [6,7] The gap opening can be further enhanced if the interface atomic bonds are soft, as shown by computation of the vibrational modes of a 1-D superlattice (Fig. 3) using nearest neighbor interactions.

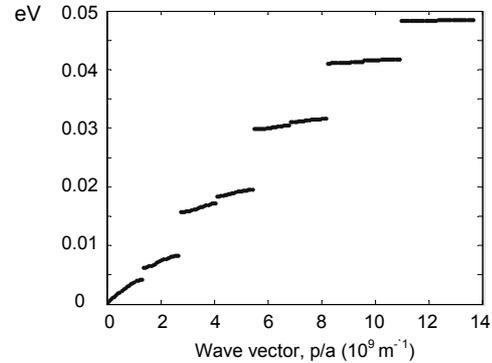

**Figure 3:** dispersion curve of acoustic phonons of a superlattice 1x1 nm when the interface modes are soft.

This result is very promising, but the gaps are sensitive to superlattice period and to acoustic impedance mismatch. Their sizes decrease rapidly when the superlattice period increases or the mismatch decreases. Furthermore the existence of such gaps is postulated on the basis of 1D models, so it would be very useful to do a full 3D simulation.

## 4 Summary and perspectives

•Existing nanostructures, albeit at high concentrations, already have the potential to achieve efficiencies >50%.
•Controlling the optical phonon generation and decay rates is critical in controlling carrier cooling. Nanostructures seem to help very significantly in this.
•A further reduced cooling rate would very likely require quantum dots rather than wells, especially if the LO to LA phonon decay is to be reduced.
•A soft interface could help enhance phononic gaps.

---
[2] At this band gap the efficiency at max. concentration is ~56%.